\documentclass[
aps,
prb,
superscriptaddress,
twocolumn,
floatfix,
10pt
]{revtex4-2}

\usepackage[utf8]{inputenc}
\usepackage{amsmath,amssymb,mathtools}
\usepackage{bm}
\usepackage{graphicx}
\usepackage[dvipsnames]{xcolor}
\usepackage{hyperref}
\usepackage[capitalise]{cleveref}
\usepackage{physics}
\usepackage{bbm}
\usepackage{orcidlink}
\usepackage{booktabs}

\newcommand{\BZ}{\mathrm{BZ}}

\newcommand{\cX}{\mathcal X}
\newcommand{\bk}{\mathbf{k}}
\newcommand{\bq}{\mathbf{q}}
\newcommand{\bQ}{\mathbf{Q}}
\newcommand{\bR}{\mathbf{R}}
\newcommand{\br}{\mathbf{r}}

\begin{document}

\title{Topology Obstructs Nodeless Excitons}

\author{Lumen Eek}
\affiliation{
Institute of Theoretical Physics,
Utrecht University,
3584 CC Utrecht,
The Netherlands
}

\date{\today}

\begin{abstract}
Lowest-energy excitons are usually expected to have nodeless \(1s\)-like envelope wavefunctions. We show that band topology can obstruct this expectation. The exciton envelope wavefunction is a section of the valence–conduction transition bundle, whose topology can enforce momentum-space zeros, with their parity or signed total index fixed by the corresponding characteristic class. We exemplify this behavior in a minimal \(PT\)-symmetric three-band model with a rank-two transition bundle characterized by Euler number \(\nu^e=2\). By directly solving the Bethe–Salpeter equation with both contact and Rytova–Keldysh interactions, we find that the lowest exciton contains two zeros, each with index \(+1\), and is less strongly bound than its topologically trivial counterpart. We further classify topology-enforced exciton nodal structures for real and complex transition bundles in dimensions \(d\leq3\). Our results establish the topology of the transition bundle as a direct constraint on exciton wavefunctions and spectra.
\end{abstract}

\maketitle
\textcolor{SkyBlue}{\textit{Introduction---}}Excitons govern the optical response of many insulating and semiconducting materials \cite{Onida2002}. In conventional semiconductors, the lowest bound exciton is expected to have a nodeless $1s$-like envelope, while nodes are associated with higher-Rydberg or angular momentum states. Whether this no-node expectation survives when the underlying Bloch bands carry nontrivial topology is not obvious. To make this question precise, consider an exciton with center-of-mass momentum $\bQ$ can be written as,
\begin{equation}
\ket{X_{\bQ}}
=
\sum_{\bk}
\mathcal{A}^X(\bk;\bQ)
c_{c,\bk+\bQ}^{\dagger}c_{v,\bk}
\ket{\mathrm{GS}}, \notag
\end{equation}
where $c^\dagger_{i,\bk}$ is the electronic creation operator in band $i$ at momentum $\bk$ and $\ket{\mathrm{GS}}$ is the ground state, i.e. the filled Fermi sea. Within the two-band Wannier description, the internal dynamics of the electron–hole pair are treated as a scalar problem. Neglecting the momentum-space texture of the periodic Bloch functions, the envelope wavefunction obeys the Wannier equation \cite{Wannier1937}
\begin{align}
&\left[
\varepsilon_c(\bk+\bQ)-\varepsilon_v(\bk)
\right]\mathcal{A}^X(\bk;\bQ) \label{eq:scalar-wannier}\\
&-
\int_{\BZ}d^2k'\,
V(\bk-\bk')\mathcal{A}^X(\bk';\bQ)
=
E_X(\bQ)\mathcal{A}^X(\bk;\bQ). \notag
\end{align}
Here, $V(\bk-\bk')>0$ is the attractive direct interaction. If the integral kernel is irreducible and positivity improving, like the Coulomb interaction, its lowest eigenstate is real and strictly positive. In a rotationally invariant continuum, this is the familiar $1s$ exciton. The conventional scalar exciton problem therefore obeys a Perron--Frobenius-type no--node theorem \cite{ReedSimon1978}.

Bloch electrons, however, are not scalar particles: Their wavefunctions carry a momentum-dependent orbital texture. This is incorporated by projecting the electron-hole interaction on the relevant conduction and valence bands such that the bare interaction potential $V(\bk-\bk')$ should be replaced by the kernel \cite{RohlfingLouie2000, Onida2002}
\begin{equation}
    K_\bQ(\bk,\bk') = V(\bk-\bk') \langle u_{c,\bk+\bQ}|u_{c,\bk'+\bQ}\rangle \langle u_{v,\bk'}|u_{v,\bk}\rangle.
\end{equation}
Through the Bloch function overlaps, Berry curvature and quantum geometry modify binding energies, exciton wavefunctions, and optical selection rules. A striking example of this occurs in transition-metal dichalcogenide (TMD) monolayers \cite{Wang2018Colloquium}, where Berry curvature lifts the degeneracy between $p_+$ and $p_-$ excitons within a given valley \cite{SrivastavaImamoglu2015, ZhouEtAl2015}, while band winding produces unconventional optical selection rules \cite{Cao2018Unifying, Beaulieu2024BerryCurvature}. 

Beyond spectral properties, topology can constrain the global structure of the exciton envelope itself. For a complex transition line bundle, a mismatch between the conduction- and valence-band Chern numbers forces vortices in the envelope whose total winding is fixed by their Chern number difference~\cite{Xie2024}. Related momentum-space textures occur in chiral excitonic-insulator order~\cite{Wang2019TopologicalExcitonicInsulator} and in intervalley-coherent exciton vortex lattices~\cite{Bultinck2020Mechanism}.

In this work, we show that band topology can have far more drastic consequences for excitons by obstructing a nodeless envelope wavefunction. We first connect this mechanism to the Chern-enforced vorticity of excitons in complex transition bundles \cite{Bultinck2020Mechanism, Xie2024}. We then turn to $PT$-symmetric systems, where the transition bundle is real. In a minimal three-band model, a nonzero Euler class forces zeros in every exciton wavefunction; direct Bethe-Salpeter equation (BSE) solutions with contact and Rytova–Keldysh (RK) interactions demonstrate this obstruction and its effect on exciton binding. Finally, we extend the Euler construction to a general Stiefel–Whitney (SW) classification of topology-enforced exciton nodes.

\textcolor{SkyBlue}{\textit{Excitons in topological bands---}}To describe excitons in topological bands, we consider electrons in the conduction subspace $\mathcal{E}_c(\bk+\bQ)$ and holes in the dual of the valence subspace $\mathcal{E}^*_v(\bk)$. An exciton with center-of-mass momentum $\bQ$ then takes the form
\begin{equation}
\ket{X_{\bQ}}
=
\sum_{\bk,\alpha,\beta}
\mathcal{A}^X_{\alpha \beta}(\bk;\bQ)
c^\dagger_{c_\alpha,\bk+\bQ}
c_{v_\beta,\bk}^{}
\ket{\mathrm{GS}},
\label{eq:exciton-state}
\end{equation}
where $\alpha$ and $\beta$ label conduction and valence bands, respectively.
The wavefunction $\mathcal{A}^X_{\alpha \beta}$ is obtained by solving the BSE, as described in the End Matter. At each internal momentum $\bk$, the electron lies in the conduction subspace $\mathcal{E}_c(\bk+\bQ)$, whereas the hole transforms in the dual of the valence subspace, $\mathcal{E}_v^*(\bk)$. The electron--hole transition space is therefore
\begin{equation}
\cX_{\bQ}(\bk)
=
\mathcal{E}_c(\bk+\bQ)
\otimes
\mathcal{E}_v^*(\bk).
\label{eq:transition-bundle}
\end{equation}
The gauge structure of the transition space is straightforwardly obtained by considering the gauge structures of the conduction and valence subspaces,
\begin{align}
\ket{u_c(\bk+\bQ)}
&\rightarrow
\ket{u_c(\bk+\bQ)}U_c(\bk+\bQ),\notag \\
\ket{u_v(\bk)}
&\rightarrow
\ket{u_v(\bk)}U_v(\bk).
\end{align}
Consequently, the exciton wavefunction transforms as
\begin{equation}
\mathcal{A}^X(\bk;\bQ)
\rightarrow
U_c^\dagger(\bk+\bQ)
\mathcal{A}^X(\bk;\bQ)
U_v(\bk).
\label{eq:exciton-gauge-transformation}
\end{equation}
Thus, the individual components of $\mathcal{A}^X$ are not globally defined scalar
functions. The projected BSE kernel contains overlaps of Bloch states at
different momenta, which connect the corresponding electron--hole transition
spaces in a gauge-covariant way. Band geometry therefore enters both the
interaction kernel and the global structure of its exciton eigenstates.

For a complex transition bundle, this geometry leads directly to a topological constraint. Its Berry curvature is $\Omega_{\cX_{\bQ}}(\bk)=\Omega_c(\bk+\bQ)-\Omega_v(\bk)$, and hence
\begin{equation}
C_{\cX}
=
\frac{1}{2\pi}
\int_{\BZ}d^2k\,
\Omega_{\cX_{\bQ}}(\bk) =  C_c - C_v.
\label{eq:transition-chern}
\end{equation}
If $C_\cX \neq0$, the transition bundle does not allow for a globally smooth, nowhere-vanishing exciton wavefunction: \(\mathcal{A}^X(\bk)\) cannot remain nonzero throughout the Brillouin zone (BZ): it must vanish at isolated momenta \(\bk_i\). Around each zero, its phase winds by an integer multiple of \(2\pi\),
\[
\nu_i
=
\frac{1}{2\pi}
\oint_{\gamma_i}
d\arg \mathcal{A}^X,
\]where \(\gamma_i\) is a small loop enclosing \(\bk_i\). The net vorticity is fixed by the difference between the band Chern numbers
\begin{equation}
\sum_i \nu_i = C_v-C_c.
\label{eq:chern-zero-counting}
\end{equation}
Topology fixes only this total winding: the number, positions, and multiplicities of the individual vortices depend on the microscopic details of the model. The sign is opposite to the Chern number $C_{\mathcal X}$ of the transition space because $\mathcal{A}^X$ is its expansion coefficient and therefore transforms with the inverse gauge phase.

This constraint underlies the momentum-space vortices found by Xie
\textit{et al.}~\cite{Xie2024}. In the flattened Haldane model studied there,
$C_v=-C_c=1$, and the lowest exciton carries total vorticity two, with an
approximately $d+id$ momentum-space profile. The corresponding optical
selection rules were analyzed in Ref.~\cite{Lozano2025}.

\textcolor{SkyBlue}{\textit{Euler obstruction to nodeless excitons---}}We now turn to the analogous obstruction for a real transition space. Let a
two-dimensional spinless system have a momentum-local antiunitary symmetry
$\Theta=PT$ with $\Theta^2=+1$. The electronic Hamiltonian and the BSE can
then be represented locally by real matrices. If the relevant transition
bundle $\cX_{\bQ}$ is a real bundle of rank two, a nondegenerate
exciton eigenstate can be chosen to have a real wavefunction. Choosing a smooth orthonormal basis of transition
states $\{\ket{\tau_1(\bk;\bQ)},\ket{\tau_2(\bk;\bQ)}\}$. The
momentum-resolved exciton amplitude can be expanded as
\begin{equation}
\ket{X_{\bQ}(\bk)}
=
A_1^X(\bk;\bQ)\ket{\tau_1(\bk;\bQ)}
+
A_2^X(\bk;\bQ)\ket{\tau_2(\bk;\bQ)},
\label{eq:real-exciton-basis}
\end{equation}
$A_1^X,A_2^X\in\mathbb{R}$. A zero occurs when both real components vanish. If $\bk_i$ is an isolated
zero, its index is the winding of the two-component amplitude around a small
counterclockwise loop $\gamma_i$,
\begin{equation}
\nu_i
=
\frac{1}{2\pi}
\oint_{\gamma_i}
d\arg\!\left(A_1^X+iA_2^X\right).
\label{eq:local-zero-index}
\end{equation}

Different local transition bases are related by momentum-dependent $SO(2)$
rotations. The associated Berry connection and curvature of the real
transition doublet are
\begin{equation}
a_i^{\cX}(\bk)
=
\left\langle
\tau_1(\bk;\bQ)
\middle|
\partial_{k_i} \middle|\tau_2(\bk;\bQ)
\right\rangle,
\quad
\mathcal{F}_{xy}^{\cX}
=
\partial_{k_x}a_y^{\cX}
-
\partial_{k_y}a_x^{\cX}.
\label{eq:euler-connection-curvature}
\end{equation}
Under a basis rotation by an angle $\phi(\bk)$,
\begin{equation}
A_1^X+iA_2^X
\rightarrow
e^{-i\phi(\bk)}
\left(A_1^X+iA_2^X\right),
\quad
\mathbf{a}^{\cX}
\rightarrow
\mathbf{a}^{\cX}-\nabla_{\bk}\phi.
\end{equation}
Although the phase of $A_1^X+iA_2^X$ depends on the local basis, the index of
an isolated zero is unchanged by any smooth basis change. For isolated zeros, their total index is fixed by the
integrated curvature,
\begin{equation}
\sum_i\nu_i
=
\frac{1}{2\pi}
\int_{\BZ}d^2k\,
\mathcal{F}_{xy}^{\cX}(\bk)
\equiv
\langle e(\mathcal{X}_\bQ), [\mathbb{T}^2]\rangle = \nu^e.
\label{eq:euler-zero-counting}
\end{equation}
Here, $e(\cX_{\bQ}) \in H^2(\mathbb{T}^2; \mathbb{Z})$ is the Euler class of the transition bundle and $\nu^e$ is its Euler number. The latter is fixed by the electronic transition space, whereas the positions and individual indices of the zeros specify how a particular exciton eigenstate realizes this topology.

A nonzero Euler number therefore prevents any real exciton wavefunction from being nonzero everywhere yielding \textit{Euler-obstructed excitons}. This constraint applies to every exciton eigenstate represented by a real wavefunction, not only to the lowest bound state. The lowest state is nevertheless of particular interest: for a scalar, positivity-improving Wannier kernel, it would be nodeless. For a degenerate exciton multiplet, one can choose a real eigenbasis, although generic complex linear combinations need not satisfy the constraint for an individual real wavefunction.

\begin{figure}
    \centering
    \includegraphics[width=\linewidth]{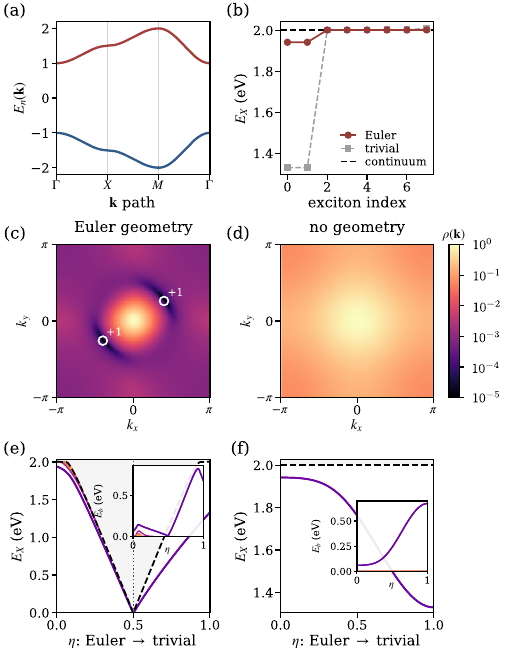}
    \caption{\textit{Euler obstruction to a nodeless exciton.}
(a) Electronic band structure of the three-band model with
$\deg(\mathbf{n})=1$. (b) Exciton spectrum obtained from the BSE with an
onsite attraction $V(\bq)=U$, compared with that of a topologically trivial
reference model having the same electronic dispersion. (c,d)
Internal-momentum probability density of the lowest bound exciton in the
Euler-nontrivial and trivial models, respectively. The
Euler-nontrivial exciton has two isolated zeros, each with index $+1$, as
required by $\nu^e=2$, whereas the corresponding exciton in the trivial
model is nodeless.}
    \label{fig:EulerExcitons}
\end{figure}

\textcolor{SkyBlue}{\textit{Minimal three--band model---}}We illustrate this mechanism with a real three-band model consisting of one conduction band and an Euler-nontrivial two-band valence subspace. The
conduction eigenstate is specified by the unit vector $\mathbf{n}(\bk)$, and the valence subspace is the plane orthogonal to $\mathbf{n}(\bk)$. The Bloch Hamiltonian reads \footnote{This particular choice of the Bloch Hamiltonian allows us to separate the effects of band geometry and band dispersion. In the End Matter we present additional models where they are more physically related.}
\begin{equation}
H(\bk)
=
D(\bk)
\left[
2\mathbf{n}(\bk)\mathbf{n}(\bk)^T-\mathbb{I}_3
\right].
\label{eq:three-band-hamiltonian}
\end{equation}
Here, $D(\bk) \in \mathbb R$ sets the band dispersion, such that the spectrum is given by $E_n(\bk) = \{ -D(\bk), -D(\bk), +D(\bk) \} $, see Fig~\ref{fig:EulerExcitons}(a). The valence subspace is the pullback of the tangent bundle of $S^2$ under the map $\mathbf{n}:\mathbb{T}^2\rightarrow S^2$. With the orientation induced by $\mathbf{n}$, its Euler number is
\begin{align}
\nu^e
&=
\frac{1}{2\pi}
\int_{\BZ}d^2k\,
\mathbf{n}(\bk)\cdot
\left[
\partial_{k_x}\mathbf{n}(\bk)
\times
\partial_{k_y}\mathbf{n}(\bk)
\right]
\nonumber\\
&=
2\,\deg(\mathbf{n}).
\label{eq:euler-degree}
\end{align}
Hence, in this case the Euler number of the valence subspace is twice the degree of $\mathbf{n}$. Because the conduction bundle is trivial, the full transition bundle has Euler number $\nu^e$. 

We now demonstrate this obstruction in the simplest setting. We take $D(\bk) = 1+(2-\cos k_x - \cos k_y)/4$, which yields a direct gap of $2~\text{eV}$ at the $\Gamma$ point. We choose $\mathbf{n}(\bk)=\mathbf{d}(\bk)/|\mathbf{d}(\bk)|$, with $\mathbf{d}(\bk) = (\sin k_x, \sin k_y, -1+\cos k_x+\cos k_y)^T$. This choice of $\mathbf{n}$ has degree $1$ such that $\nu^e=2$. We first solve the BSE while considering only an onsite electron--hole interaction, $V(\bq)=U=3/2$. Figure~\ref{fig:EulerExcitons}(b) shows the resulting exciton energies $E_X$. The spectrum contains two degenerate bound excitons, reflecting the two interband transition channels from the two valence bands to the conduction band, each of which supports one bound state for a contact interaction. Their degeneracy is enforced by additional crystalline symmetries of the model. Breaking these symmetries splits the two exciton energies but leaves their Euler obstruction intact. In Fig.~\ref{fig:EulerExcitons}(c), we depict the (normalized) momentum-space exciton density $\rho(\bk) \equiv \sum_{\alpha\beta} |A_{\alpha\beta}^X (\bk)|^2$ of one of the bound excitons (the other one is related by a $C_4$ rotation). The density shows two zeros, indicated by the white circles, that both contribute $+1$ to the total winding. We compare this result to one obtained from a Hamiltonian with the exact same dispersion and interaction, but which is topologically trivial, with $\mathbf{n}(\bk) = (1,0,0)^T$. The corresponding exciton energies and the lowest-state momentum-space density are shown in Figs.~\ref{fig:EulerExcitons}(b) and (d), respectively. Importantly, there is no nodes in Fig.~\ref{fig:EulerExcitons}(d), which is corroborated by the trivial Euler number, $\nu^e=0$.

The zeros in the exciton wavefunction can only be removed by closing the single-particle gap or by breaking $PT$ symmetry. We illustrate the former in Fig.~\ref{fig:EulerExcitons}(e), using the interpolating Bloch Hamiltonian $H(\eta) \equiv (1-\eta )H_{\text{Euler}} + \eta H_\text{triv}$. The colored lines correspond to bound exciton energies, while the dashed black line denotes the single-particle gap. The inset shows the corresponding exciton binding energies $E_b = E_\text{gap}-E_X$. As $\eta$ increases from $0$ to $1$, the system goes through a gap-closing around $\eta = 1/2$ to end up in the trivial phase. Notably, additional bound states appear on the topological side of the transition, indicating that the Bloch form factors can sufficiently reshape the interaction Kernel to support additional bound states. Further details are provided in the Supplemental Material. Additionally, we consider a second path in parameter space where we keep the gap open and instead interpolate $\mathbf{n}$ (see End Matter for details). The results are shown in Fig.~\ref{fig:EulerExcitons}(f). In this case $PT$ symmetry is broken along the path except for the end points.

\textcolor{SkyBlue}{\textit{Topological constraints on exciton binding---}}In addition to enforcing zeros in the exciton wavefunction, the Euler obstruction modifies the exciton energy, as seen in Fig.~\ref{fig:EulerExcitons}(b). In a scalar attractive problem, the lowest-state wavefunction can align its sign throughout momentum space, thereby maximizing the attractive interaction matrix element. A nontrivial transition bundle prevents such a globally uniform alignment. The required zeros and their associated winding consequently reduce the interaction energy gained upon binding.

The magnitude of this effect is not topologically quantized. Instead, it depends on the momentum-space extent of the exciton and on the distribution of Euler curvature across the BZ. A weakly bound Wannier exciton is concentrated near the band edge and samples only a small region of the BZ which could carry little Euler curvature. Within this region, the transition bundle is locally trivial, and the exciton wavefunction and energy can remain close to the topologically trivial limit. In this case, the exciton wavefunction will still have zeros, but they may lie in regions of negligible exciton weight. The Euler obstruction is therefore most consequential for tightly-bound excitons, whose broad momentum-space wavefunctions sample a substantial fraction of the BZ.

\begin{figure}
    \centering
    \includegraphics[width=\linewidth]{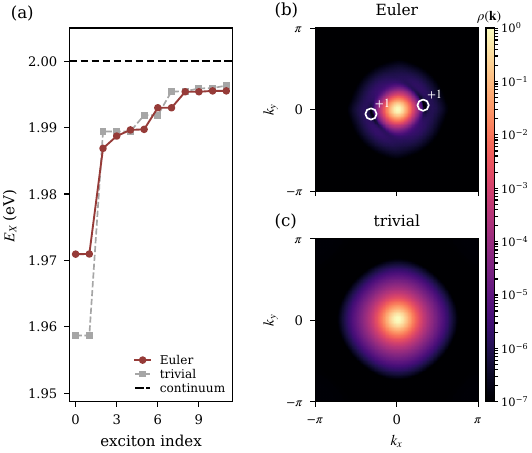}
    \caption{\textit{Euler excitons with a RK interaction.}
(a) Exciton spectra for the Euler and topologically trivial band geometries, with identical single-particle dispersions [given in Fig.~\ref{fig:EulerExcitons}(a)].
(b),(c) Momentum-space densities of the lowest exciton for the Euler and trivial geometries, respectively. The white circles in (b) indicate the zeros. We take $g=0.25~\text{eV}$ and $r_0 = a$, with $a$ the lattice constant.}
    \label{fig:RK}
\end{figure}

\textcolor{SkyBlue}{\textit{Euler excitons from realistic interaction potentials---}}The contact interaction considered above makes the topological behavior particularly visible because the corresponding Frenkel-like excitons sample the whole BZ. Now, we will demonstrate that the obstruction is also present upon solving the BSE with a more realistic interaction potential. To this end, we consider the RK potential $V(\mathbf{r}) = g \left[ H_0(|\mathbf{r}|/r_0) - Y_0(|\mathbf{r}|/r_0) \right]$ \cite{Rytova1967, Keldysh1979, Chernikov2014Nonhydrogenic}. Here, $g$ sets the interaction strength, $r_0$ is the screening length, and $H_0$ and $Y_0$ are the Struve function of order zero and the zeroth-order Bessel function of the second kind, respectively. Figure~\ref{fig:RK}(a) shows the resulting exciton spectra for the Euler and trivial band geometries, using the same electronic dispersions as in Fig.~\ref{fig:EulerExcitons}(a). The lowest Euler exciton is less strongly bound than its topologically trivial counterpart. Figures~\ref{fig:RK}(b) and \ref{fig:RK}(c) show the momentum-space densities of the corresponding lowest exciton states. While the trivial exciton exhibits a conventional $1s$-like profile, the Euler exciton contains two zeros, each carrying Euler charge $+1$, as indicated by the white circles.

The comparison with the contact-interaction limit illustrates the distinction between topological and spectral information. The RK interaction reshapes the exciton wavefunction and changes the bound-state energies, while the two zeros, with total Euler charge $+2$, persist. More generally, changing the interaction can move, split, or merge the zeros and rearrange the exciton levels, but cannot change their total index without closing the single-particle gap or breaking $PT$ symmetry. The Euler number therefore fixes the total nodal charge of the exciton wavefunction, but does not determine its binding energy or the ordering of the exciton levels.
\begin{table}[t]
\caption{\label{tab:real-zero-classification}%
Topology-enforced zeros of a real exciton wavefunction
$\mathcal{A}^X$. For each spatial dimension $d$ and transition-bundle
rank $r$, we list the relevant SW invariants for both
orientable and nonorientable $\mathcal{X}_{\mathbf Q}$, together with
the relations constraining the lower SW numbers. When $\mathcal{X}_{\mathbf Q}$ is orientable, we express the SW invariant in terms of the Euler invariant $\nu_{\Sigma_r}^{e}$. A dagger marks an orientable odd-rank case in which the Euler class exists but must vanish on the Brillouin torus \cite{milnor-stasheff}. A dash indicates that no invariant of the specified degree is available because $i>r$ or $i>d$, while ``any'' means that the corresponding invariant is unconstrained by the relations shown. 
}
\centering
\scriptsize
\setlength{\tabcolsep}{3.5pt}
\renewcommand{\arraystretch}{1.15}

\begin{ruledtabular}
\begin{tabular}{cc|ccc|l}
$d$ & $r$ & $w_1$ & $w_2$ & $w_3$
    & Enforced zeros \\
\hline
$1$ & $1$ & $0^\dagger$     & --          & --    & none \\
    & $1$ & $ \nu^{w_1}_{\mathbb{T}}$ & --          & --    & points ($0$D) \\[2pt]

$2$ & $1$ & $0$     &  --         & --    & none \\
    & $1$ & $ \boldsymbol{\nu}^{w_1}$ & --         & --    & curves ($1$D) \\
    & $2$ & $0$     & $\nu^{e}_{\mathbb{T}^2}\bmod 2$  & --    & points ($0$D) \\
    & $2$ & $\ne 0$ & $\nu^{w_2}_{\mathbb{T}^2}$       & --    & points ($0$D) \\[2pt]

$3$ & $1$ & $0$     & --        & --   & none \\
    & $1$ & $\boldsymbol{\nu}^{w_1}$ & --        & --   & surfaces ($2$D) \\
    & $2$ & $0$     & $\boldsymbol{\nu}^{e}\bmod 2$  & --  & curves ($1$D) \\
    & $2$ & $\ne 0$ & $\boldsymbol{\nu}^{w_2}$       & --   & curves ($1$D) \\
    & $3$ & $0$     & any         & $0^\dagger$   & none \\
    & $3$ & $\ne 0$ & any         & $\nu^{w_3}_{\mathbb{T}^3}$ & points ($0$D) \\
\end{tabular}
\end{ruledtabular}
\end{table}

\textcolor{SkyBlue}{\textit{Classification---}}The three-band model introduced above is one example of a broader class of models in which the exciton envelope wavefunctions are constrained by the topology of the transition bundle. This behavior can be extended to systems with transition bundles of rank $r$ in $d$ spatial dimensions by a counting argument. At each momentum, the exciton wavefunction consists of $N= n_\mathbb{F}r$ real components, where $n_\mathbb{F}=n_\mathbb{R}=1$ ($n_\mathbb{C}=2$) for real (complex) transition bundles. Therefore, a zero occurs when all $N$ components vanish simultaneously. This yields as the dimension of the nodal set
\begin{equation}
    d_\text{node} = d-n_\mathbb{F}r.
\end{equation}
Consequently, $d=n_\mathbb{F} r$ yields isolated point nodes, while $d=n_\mathbb{F} r+1$ and $d=n_\mathbb{F} r+2$ give nodal lines and nodal surfaces, respectively. In the three-band model considered above, $d=2$, $r=2$, and $\mathbb F=\mathbb R$, such that the topology-enforced zeros are isolated points.

This argument specifies the dimension of the zero set, but does not specify whether nodes must exist. For a real rank-$r$ transition bundle, the relevant invariant is given by its top SW class $w_r(\mathcal{X}_\bQ) \in H^r(\mathbb{T}^d;\mathbb{Z}_2)$. Evaluating the SW class on a $r$-dimensional Brillouin-zone cycle $\Sigma_r$ yields the SW invariant $\nu_{\Sigma_r}^{w_r} \equiv \langle w_r(\mathcal{X}_\bQ), [\Sigma_r]\rangle \in \mathbb{Z}_2$. The cycles $\Sigma_r$ of $\mathbb{T}^d$ can be generated by the $\binom{d}{r}$ subtori $\mathbb{T}^{r}_I$. For $d=2$, the independent one-cycles are $\mathbb T_x$ and $\mathbb T_y$, and we define
\[
\boldsymbol{\nu}^{w_1}
=
\left(
\nu_{\mathbb T_x}^{w_1},
\nu_{\mathbb T_y}^{w_1}
\right)
\in(\mathbb Z_2)^2.
\]
For \(d=3\), the independent one- and two-cycles are
$
\Sigma_1\in
\left\{
\mathbb T_x,\mathbb T_y,\mathbb T_z
\right\}$,
$\Sigma_2\in
\left\{
\mathbb T_{xy}^2,\mathbb T_{yz}^2,\mathbb T_{zx}^2
\right\}$,
and the corresponding invariants are
\[
\boldsymbol{\nu}^{w_i}
=
\left(
\nu_{\mathbb T_x}^{w_i},
\nu_{\mathbb T_y}^{w_i},
\nu_{\mathbb T_z}^{w_i}
\right)
\in(\mathbb Z_2)^3 \qquad i=1,2.
\]
A top nonzero \(\nu_{\Sigma_r}^{w_r}\) obstructs a nowhere-vanishing exciton wavefunction on \(\Sigma_r\), and therefore enforces a nodal set in the BZ. The \(w_1\) and \(w_2\) invariants may be obtained using Wilson-loop methods \cite{Ahn2018BandTopology, Ahn2019StiefelWhitney}. 

If $w_1(\mathcal{X}_\bQ) =0$, i.e. all $\nu^{w_1}_{\Sigma_1} =0$, then $\mathcal{X}_\bQ$ is orientable and admits an Euler class $e(\cX_\bQ) \in H^r(\mathbb{T}^d; \mathbb{Z})$. The Euler class is the reduction modulo two of the top SW class: $e(\cX_\bQ)  \mod 2= w_r(\cX_\bQ)$, i.e., for a generic exciton wavefunction, the SW invariant gives the parity of its zeros, while the Euler invariant gives the sum. When $d=r$ the Euler class is evaluated over the full BZ to yield the Euler number ${\nu^e_{\mathbb{T}^d} \equiv\langle e(\cX_\bQ), [\mathbb{T}^d] \rangle}$, as introduced in Eq.~\eqref{eq:euler-zero-counting}. It is worth noting that on the torus, the Euler number is only non-zero for $d=2$ \cite{milnor-stasheff}, such that in physical systems it is only non-zero in two dimensions. However, in $d=3$ a rank-$2$ Euler class can still be evaluated over two-dimensional cycles of the BZ, yielding $\nu_{\mathbb{T}_{ij}}^{w_2} = \nu_{\mathbb{T}_{ij}}^{e}\mod2$. Euler-obstructed excitons therefore constitute the orientable, integer refinement of \textit{Stiefel-Whitney-obstructed excitons}, i.e. real-bundle-obstructed excitons. When $\nu^e_{\Sigma_r}$ is odd, the Euler obstruction is also detected by $\nu^{w_r}_{\Sigma_r}$. In Table~\ref{tab:real-zero-classification}, we summarize these constraints on the exciton wavefunctions for all relevant $r \leq d\leq3$.

Applying the same procedure to complex transition bundles yields the corresponding classification of Chern-obstructed excitons in terms of Chern classes, summarized in Table~\ref{tab:complex-zero-classification}. The classification is considerably more restricted because a complex rank-$r$ exciton wavefunction has $2r$ real components. For $d=2$ and $r=1$, a nonzero $C_1$ enforces point vortices; this is the case studied by Xie \textit{et al.} \cite{Xie2024}. In $d=3$, the same $c_1$ obstruction instead produces nodal lines. Higher Chern classes enter only in higher dimensions. The first such case is $d=4$ and $r=2$, where a nontrivial $c_2$ enforces isolated point nodes.

It is important to realize that the Euler obstruction is fragile already at the single-particle level; Consider the Euler class over a rank two bundle in two spatial dimensions. Adding additional (trivial) bands reduces the $\mathbb Z$-valued Euler classification back to the stable $\mathbb Z_2$ SW classification. Consequently, an even Euler number, such as $\nu^e_{\mathbb{T}^2}=2$ in our present model, can be trivialized by adding trivial channels \cite{Ahn2019Failure}. At the exciton level; once the rank of the transition bundle exceeds the spatial dimension, isolated zeros of a generic exciton wavefunction are no longer enforced by dimensional counting. The nodal structure of Euler-obstructed excitons is therefore protected within an isolated fixed-rank transition subspace, such that it itself is also fragile to the inclusion and hybridization of additional transition channels. This fixed-rank fragility of the nodal structure applies equally to Euler-, SW-, and Chern-obstructed excitons, even when a stable lower characteristic class of the enlarged transition bundle remains nontrivial.

\begin{table}[t]
\caption{\label{tab:complex-zero-classification}%
Topology-enforced zeros of a complex exciton wavefunction
$\mathcal{A}^X$. For each spatial dimension $d$ and transition-bundle
rank $r$, we list the relevant Chern numbers. A dash indicates that no invariant of the specified degree is available ($c_2$ only becomes available in four spatial dimensions). $\mathrm{C}_1$ denotes the first Chern number evaluated over the full two-dimensional BZ, while $\mathbf{C}_1$ denotes the vector of first Chern numbers evaluated over the two-dimensional cycles of the three-dimensional BZ.
}
\centering
\scriptsize
\setlength{\tabcolsep}{5pt}
\renewcommand{\arraystretch}{1.15}

\begin{ruledtabular}
\begin{tabular}{ccccl}
$d$ & $r$ & $c_1$ & $c_2$ & Enforced zeros \\
\hline
$1$ & $1$ & --      & --      & none \\[2pt]

$2$    & $1$ & $\mathrm{C}_1$ & --      & points ($0$D) \\
    & $2$ & any     & --      & none \\[2pt]

$3$    & $1$ & $\mathbf{C}_1$ & --      & curves ($1$D) \\
    & $2$ & any     & --      & none \\[2pt]

\end{tabular}
\end{ruledtabular}
\end{table}

\textcolor{SkyBlue}{\textit{Discussion \& Conclusions---}}In this work, we identified Stiefel--Whitney- and Euler-obstructed excitons, whose envelope wavefunctions are sections of topologically nontrivial real transition bundles. For a general real transition bundle, the top SW class detects the mod-two obstruction to a nodeless exciton wavefunction. When the bundle is orientable, the Euler class provides an integer refinement that can remain nonzero even when the SW invariant vanishes. For isolated zeros, the SW invariant fixes their parity, whereas the Euler invariant fixes their signed total index.

We demonstrated this obstruction in a minimal $PT$-symmetric three-band model with a rank-two transition bundle of Euler number $\nu^e_{\mathbb{T}^2}=2$. Solving the BSE with both contact and RK interactions, we found that the lowest exciton exhibits two momentum-space zeros, each with index $+1$, whereas its topologically trivial counterpart has a conventional nodeless $1s$-like profile. In our model, the Euler exciton is also less strongly bound than its topologically trivial counterpart. The magnitude of this reduction is not quantized and depends on the interaction range, band dispersion, and momentum-space extent of the exciton. We therefore expect the energetic signature of the obstruction to be most pronounced for tightly bound excitons.

The topology-enforced zeros identified here are distinct from the radial or angular nodes of excited hydrogenic states and from the symmetry-enforced exciton zeros studied by Hwang \textit{et al.} \cite{Hwang2026}. The latter case investigates exciton zeros at high-symmetry momenta, enforced by inversion and rotation eigenvalues. Here, by contrast, the zeros arise from a global obstruction of the transition bundle and require neither nonzero relative angular momentum, nor crystalline-symmetry pinning. Additional crystalline symmetries may constrain the positions and multiplicities of the zeros, but breaking these symmetries can move, split, or merge the zeros without changing their total index as the real structure, the electronic gap isolating the transition subspace, and the topology of the transition bundle are preserved.

Furthermore, our analysis concerns the internal momentum dependence of a single exciton at fixed center-of-mass momentum $\mathbf Q$. It is distinct from Berry-phase effects in exciton center-of-mass dynamics and transport~\cite{YaoNiu2008,Tang2024Inheritance}, as well as from the topology of exciton bands over $\mathbf Q$~\cite{ChenShindou2017,WuLovornMacDonald2017,Kwan2021ExcitonBand,Davenport2024}. Nevertheless, internal and center-of-mass topology become naturally coupled when $\mathbf Q$ is allowed to vary. The full envelope $\mathcal A^X(\mathbf k;\mathbf Q)$ can then be viewed as a family of internal wavefunctions over the product space $\mathbb T_{\mathbf k}^d\times\mathbb T_{\mathbf Q}^d$. Topology along $\mathbf k$ constrains the internal nodal texture, whereas topology along $\mathbf Q$ controls the transport and boundary properties of the exciton band. Investigating the interplay between internal $\mathbf k$ and center-of-mass $\mathbf Q$ topology remains an interesting direction for future work.

A complementary extension would be the investigation of three-dimensional band topology beyond characteristic classes in the exciton problem. In a two-band Hopf insulator, all first Chern numbers on two-dimensional Brillouin-zone cycles can vanish even though the map $\mathbf n:\mathbb T^3\rightarrow S^2$ carries a nonzero Hopf invariant~\cite{Moore2008Hopf, Deng2013Hopf, Nelson2021Delicate, Lapierre2021NBandHopf, Lim2023RealHopf, Jankowski2024HopfEuler}. The corresponding complex line bundle is topologically trivial, so the Hopf invariant does not by itself enforce zeros of a single exciton envelope in the same manner as a Chern or SW/Euler class. Nevertheless, the BSE kernel retains the full momentum-space texture of the Bloch states and may imprint the linked structure of the Hopf phase on the exciton wavefunction. 

Promising material settings are spinless, or effectively spin-decoupled, two-dimensional insulators with \(PT\) or \(C_{2z}T\) symmetry, an isolated low-rank transition subspace, and sufficiently tightly-bound excitons. Narrow-band moiré systems are a viable candidate in this respect: the nearly flat bands of single-valley twisted bilayer graphene provide a canonical realization of $C_{2z}T$-protected Euler topology \cite{Ahn2019Failure}. Weak-spin-orbit multiorbital semiconductors and two-dimensional covalent networks such as monolayer graphdiyne and liganded Xenes with nontrivial SW topology, provide another promising direction \cite{Lee2020Graphdiyne,Pan2022LigandedXenes}. Real multiband topology has also been identified in strained ZrTe \cite{Bouhon2020NonAbelian}, though it is a semimetal, making it less suitable as an excitonic platform. 

Our results shift the relevant topology from the individual electron and hole bands to the space of optically active transitions between them. In general, the topological behavior of the exciton is set by the transition bundle and can not be traced back to a single band. Its topology directly constrains the internal exciton wavefunction and leaves signatures in the exciton spectrum. Excitons therefore provide a natural probe of multiband Bloch-state topology beyond conventional single-particle diagnostics.

\textit{Acknowledgements---}I would like to thank Piet Brouwer and Rodrigo Arouca for fruitful discussions and Cristiane Morais Smith for a careful reading of the manuscript. I used AI (GPT6 and Codex) as an interactive tool to explore the topic and to structure the manuscript/code.

\bibliography{bib}

\section*{End Matter}

\subsection{Additional details on the three-band model}
\textit{Model.---}In the main text we consider 
\begin{equation}
    H(\bk) = D(\bk) [ 2\mathbf{n}(\bk) \mathbf{n}(\bk)^T - \mathbb{I}_3],
\end{equation}
with $D(\bk)>0$ and $|\mathbf{n}(\bk)|=1$. The conduction subspace is spanned by $\mathbf{n}(\bk)$, with eigenvalue $D(\bk)$ \footnote{It is straightforward to check this by evaluating $H(\bk) \mathbf{n}(\bk) = D(\bk) \mathbf{n}(\bk)$}. The degenerate valence subspace has energy $-D(\bk)$ and is formed by the two eigenvectors $\mathbf{v}_a$ that span the plane perpendicular to $\mathbf{n}(\bk)$, i.e.
\begin{equation}
    \mathbf{v}_a \cdot \mathbf{n}=0, \quad \mathbf{v}_1 \times \mathbf{v}_2 = \mathbf{n}.
\end{equation}
Here, and in the remainder of this section, we drop the explicit $\bk$-dependence of $\mathbf{v}_a$ and $\mathbf{n}$. On the valence subspace we may define the $SO(2)$ Berry connection and curvature
\begin{equation}
    a_i = \mathbf{v}_1 \cdot \partial_i \mathbf{v}_2, \quad \mathcal{F}_{ij} = \partial_i a_j - \partial_j a_i
\end{equation}
Using $\mathbf{n}\cdot \partial_i\mathbf{n} = \partial_i (\mathbf{n}\cdot \mathbf{n})/2=0$, we can write
\begin{equation}
\partial_i\mathbf n=\alpha_i\mathbf v_1+\beta_i\mathbf v_2,
\end{equation}
orthonormality gives
\begin{equation}
\partial_i\mathbf v_1=-a_i\mathbf v_2-\alpha_i\mathbf n,\qquad \partial_i\mathbf v_2=a_i\mathbf v_1-\beta_i\mathbf n.
\end{equation}
It follows that
\begin{align}
\mathcal{F}_{ij}&=(\partial_i\mathbf v_1)\cdot(\partial_j\mathbf v_2)-(i\leftrightarrow j)\nonumber\\
&=\alpha_i\beta_j-\alpha_j\beta_i\nonumber\\
&=\mathbf n\cdot\left(\partial_i\mathbf n\times\partial_j\mathbf n\right).
\end{align}
The Euler number of the valence bundle is therefore
\begin{align}
\nu^e&=\frac{1}{2\pi}\int_{\BZ}d^2k\,\mathcal{F}_{xy}\nonumber\\
&=\frac{1}{2\pi}\int_{\BZ}d^2k\,\mathbf n\cdot\left(\partial_{k_x}\mathbf n\times\partial_{k_y}\mathbf n\right)\nonumber\\
&\equiv 2\deg(\mathbf n)
\end{align}

\textit{Isospectral interpolation through broken $PT$.---}In this section, we briefly explain the construction of the interpolating Hamiltonian that preserves the single-particle spectrum but breaks $PT$ at intermediate $\eta$, the exciton spectrum along this path is shown in main text Fig.~\ref{fig:EulerExcitons}(f). Let $\mathbf n_E(\bk)$ denote the Euler conduction eigenvector and $\mathbf n_0$ the constant trivial real vector defining the trivial endpoint. We introduce
\begin{equation}
|\mathbf{n}_\eta(\bk)\rangle=\cos\left(\frac{\pi\eta}{2}\right)|\mathbf n_E(\bk)\rangle+i\sin\left(\frac{\pi\eta}{2}\right)|\mathbf n_0\rangle.
\end{equation}
Because both endpoint vectors are real, the relative factor $i$ cancels the cross terms in the norm, giving $\langle c_\eta|c_\eta\rangle=1$. We then define
\begin{equation}
\widetilde h_\eta(\bk)=D(\bk)\left[2|\mathbf{n}_\eta(\bk)\rangle\langle \mathbf{n}_\eta(\bk)|-\mathbb{I}_3\right].
\end{equation}
This interpolating Hamiltonian will leave the single-particle gap unchanged, while smoothly going from nontrivial Euler to a trivial transition bundle. Along this path $PT$ symmetry is broken, hence the system can transition without closing the gap.

\textit{Warping.---}For the RK calculations in the main text, the lowest exciton is concentrated near the $\Gamma$ point and samples only a limited region of the Brillouin zone. The Euler-enforced zeros affect its binding only substantially when they occur within a region of substantial exciton weight. To redistribute the quantum geometry while keeping both the electronic dispersion and Euler class fixed, we compose the Bloch texture with the smooth periodic map
\begin{equation}
X(k_x)=k_x+\alpha\sin k_x,\qquad Y(k_y)=k_y+\alpha\sin k_y,
\end{equation}
and replace $\mathbf n_0(k_x,k_y)$ by $\mathbf n_\alpha(\bk)=\mathbf n_0(X,Y)$, while leaving $D(\bk)$ unchanged. The map has degree one and is homotopic to the identity, so it leaves $\deg(\mathbf n)$, and hence the Euler number, unchanged. However, the curvature transforms as
\begin{equation}
\mathcal F_\alpha(\bk)=(1+\alpha\cos k_x)(1+\alpha\cos k_y)\mathcal F_0(X,Y),
\end{equation}
and is therefore redistributed across the Brillouin zone together with the quantum metric. We take $\alpha=0.9$.

\subsection{Bethe--Salpeter equation}
\textit{Band projection.---}We consider electrons subject to a translationally invariant density--density interaction,
\begin{equation}
\begin{aligned}
H_0&=\sum_{\bk}\hat{\Psi}_{\bk}^{\dagger}h(\bk)\hat{\Psi}_{\bk},\\
H_{\mathrm{int}}&=\frac{1}{2\Omega}\sum_{\bq}V(\bq):\rho_{\bq}\rho_{-\bq}:,
\end{aligned}
\end{equation}
where $\Omega$ is the system area and
\begin{equation}
\rho_{\bq}=\sum_{\bk}\hat{\Psi}_{\bk+\bq}^{\dagger}\hat{\Psi}_{\bk}.
\end{equation}
We transform to the band basis using
\begin{equation}
\begin{aligned}
\hat{\Psi}_{\bk}=U(\bk)\hat{\mathbf c}_{\bk}, \quad
U^\dagger(\bk)h(\bk)U(\bk)=\varepsilon(\bk),
\end{aligned}
\end{equation}
where $\varepsilon(\bk)=\operatorname{diag}_n[\varepsilon_n(\bk)]$. The density operator becomes
\begin{equation}
\rho_{\bq}=\sum_{\bk,m,n}\Lambda_{mn}(\bk+\bq,\bk)c^\dagger_{m,\bk+\bq}c_{n,\bk},
\end{equation}
with
\begin{equation}
\Lambda_{mn}(\bk',\bk)\equiv\langle u_{m,\bk'}|u_{n,\bk}\rangle.
\end{equation}
Projecting onto the single electron--hole sector and retaining only the direct interaction yields the direct-only BSE~\cite{SrivastavaImamoglu2015,Xie2024}
\begin{align}
&E_X(\bQ)\mathcal A^X_{cv}(\bk;\bQ)=\bigl[\varepsilon_c(\bk+\bQ)-\varepsilon_v(\bk)\bigr]\mathcal A^X_{cv}(\bk;\bQ)\nonumber\\
&-\int_{\BZ}\frac{d^d k'}{(2\pi)^d}\sum_{c'v'}K^{\mathrm d}_{\bQ;cv,c'v'}(\bk,\bk')\mathcal A^X_{c'v'}(\bk';\bQ),
\end{align}
with
\begin{equation}
\begin{aligned}
K^{\mathrm d}_{\bQ;cv,c'v'}(\bk,\bk')&=\\
V(\bk-\bk')&\Lambda_{cc'}(\bk+\bQ,\bk'+\bQ)\Lambda_{v'v}(\bk',\bk).
\end{aligned}
\end{equation}

\textit{Contact interaction.---}For an onsite electron--hole attraction, the interaction is independent of momentum transfer,
\begin{equation}
V(\bq)=U,\qquad U>0.
\end{equation}
At $\bQ=0$, we discretize the Brillouin zone using $N_p$ momenta and denote the conduction Bloch vector at momentum $\bk_i$ by $\mathbf n_i$, and the two valence Bloch vectors by $\mathbf v_{ia}$, with $a=1,2$. The direct BSE then reads
\begin{equation}
H^{\mathrm{BSE}}_{ia,jb}=\Delta_i\delta_{ij}\delta_{ab}-\frac{U}{N_p}(\mathbf n_i\cdot\mathbf n_j)(\mathbf v_{ia}\cdot\mathbf v_{jb}),
\end{equation}
where $\Delta_i=\varepsilon_c(\bk_i)-\varepsilon_v(\bk_i)$. Although the bare interaction is momentum independent, the interaction kernel is not: the conduction and valence form factors retain the full momentum dependence of the Bloch eigenvectors. In the topologically trivial control, we keep the electronic dispersions fixed but replace the Bloch eigenvectors by a momentum-independent orthonormal basis. The form factors then reduce to
\begin{equation}
\mathbf n_i\cdot\mathbf n_j=1,\qquad \mathbf v_{ia}\cdot\mathbf v_{jb}=\delta_{ab},
\end{equation}
and the BSE separates into two identical scalar contact-interaction problems.

\textit{RK interaction.---}
We measure distances in units of the lattice constant $a$ and use the RK interaction
\begin{equation}
V_{\mathrm{RK}}(\br)=g\left[\mathrm{H}_0(|\br|/r_0)-Y_0(|\br|/r_0)\right],
\end{equation}
where $g>0$ sets the interaction strength, $r_0$ is the screening length, and $\mathrm{H}_0$ and $Y_0$ are the zeroth-order Struve and Bessel functions of the second kind, respectively. Setting $a=1$, we truncate the real-space interaction at $R_c=80$. On the square lattice $\bR=(n_x,n_y)$, its regularized form is
\begin{equation}
\widetilde V(\bR)=
\begin{cases}
V_{\mathrm{RK}}(a),&\bR=0,\\
V_{\mathrm{RK}}(|\bR|),&0<|\bR|\leq R_c,\\
0,&|\bR|>R_c.
\end{cases}
\end{equation}
The momentum-space interaction entering the BSE is obtained from the lattice Fourier transform
\begin{equation}
V(\bq)=\sum_{\bR}\widetilde V(\bR)e^{-i\bq\cdot\bR}.
\end{equation}
The direct BSE Hamiltonian matrix elements are then given by
\begin{equation}
\begin{aligned}
H^{\mathrm{BSE}}_{ia,jb}={}&\Delta_i\delta_{ij}\delta_{ab}\\
&-\frac{1}{N_p}V(\bk_i-\bk_j)(\mathbf n_i\cdot\mathbf n_j)(\mathbf v_{ia}\cdot\mathbf v_{jb}).
\end{aligned}
\end{equation}
The same interaction parameters and Fourier-transform prescription are used for the Euler and trivial geometries, so their difference arises entirely from the Bloch-state form factors.

\subsection{Additional Calculations}
\textit{$w_1$-obstruction in one dimension.---}Following  Table~\ref{tab:real-zero-classification}, we require a rank-$1$ transition bundle for $d=1$ to have $w_1$-obstructed excitons. Although this suggests that a two-band model is sufficient, the SW invariants of the valence and conduction bands are equal in a two-band model such that the SW invariant of the transition bundle is always trivial. We resolve this by including a third spectator band $s$. In this case we can have 
\begin{equation}
    w_1(\mathcal{E}_c) =0, \quad w_1(\mathcal{E}_v) = w_1(\mathcal{E}_s) = 1.
\end{equation}
In this case, the full three-band bundle is trivial, while the conduction-valence transition bundle has $\nu^{w_1}=1$. We choose
\begin{equation}
\begin{aligned}
|u_{c,k}\rangle&=(1,0,0)^T,\\
|u_{v,k}\rangle&=\left(0,\cos\frac{k}{2},\sin\frac{k}{2}\right)^T,\\
|u_{s,k}\rangle&=\left(0,-\sin\frac{k}{2},\cos\frac{k}{2}\right)^T.
\end{aligned}
\end{equation}
The dispersion is defined through
\begin{equation}
\Delta(k;\lambda)=\Delta_0+t(1-\cos k)+\lambda\sin k,
\end{equation}
as
\begin{equation}
\varepsilon_{c/v}(k)= \pm\frac{\Delta(k;\lambda)}{2}, \qquad \varepsilon_s(k)=\varepsilon_v(k)-\Delta_s,
\end{equation}
where $t=0.35$ sets the dispersion, $\Delta_0 =1$ is the band gap, and $\Delta_s=1$ is the spectator-band offset. We additionally introduce a term that scales with $\lambda$ which breaks inversion symmetry $P$ (while preserving $PT$). The corresponding Hamiltonian is
\begin{equation}
h(k)=\sum_{\mu=c,v,s}\varepsilon_\mu(k)|u_{\mu,k}\rangle\langle u_{\mu,k}|.
\end{equation}
\begin{figure}
    \centering
    \includegraphics[width=\linewidth]{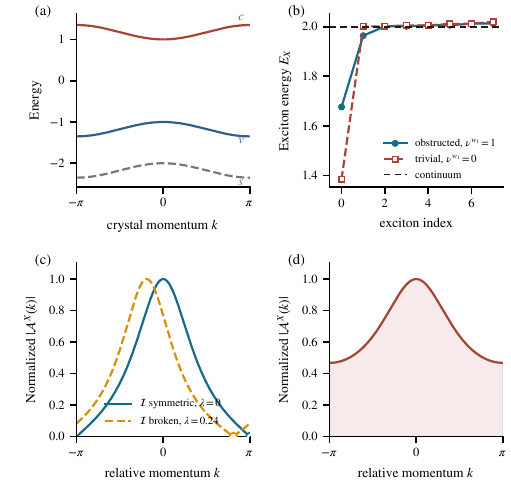}
    \caption{\textit{One-dimensional $w_1$-obstructed excitons.} (a) Single-particle spectrum of the obstructed and trivial three-band models. (b) Conduction--valence exciton spectra for a contact interaction $V(q)=U=0.8$ and $\lambda=0$. (c) Lowest obstructed exciton wavefunction for $\lambda=0$ and $\lambda=0.24$. (d) Lowest exciton wavefunction of the trivial model.}
    \label{fig:w1d1}
\end{figure}
Figure~\ref{fig:w1d1}(a) shows the single-particle spectrum, while Fig.~\ref{fig:w1d1}(b) shows the conduction--valence exciton spectrum for a contact interaction $V(q)=U=0.8$ and $\lambda=0$. Analogous to the two-dimensional results in the main text, the obstructed exciton has a smaller binding energy than its topologically trivial counterpart. In Fig.~\ref{fig:w1d1}(c), we show the lowest-energy obstructed exciton wavefunction for $\lambda=0$ and $\lambda=0.24$. Both wavefunctions contain an odd number of zeros (one in this case), as required by $\nu^{w_1}=1$. For $\lambda=0$, inversion symmetry pins this zero to the BZ boundary. A finite $\lambda$ breaks inversion symmetry and moves the zero to an arbitrary momentum without removing it. For comparison, Fig.~\ref{fig:w1d1}(d) shows the lowest exciton wavefunction of the isospectral trivial model. Its transition line has $\nu^{w_1}=0$, and the wavefunction is correspondingly nodeless.
\begin{figure}
    \centering
    \includegraphics[width=\linewidth]{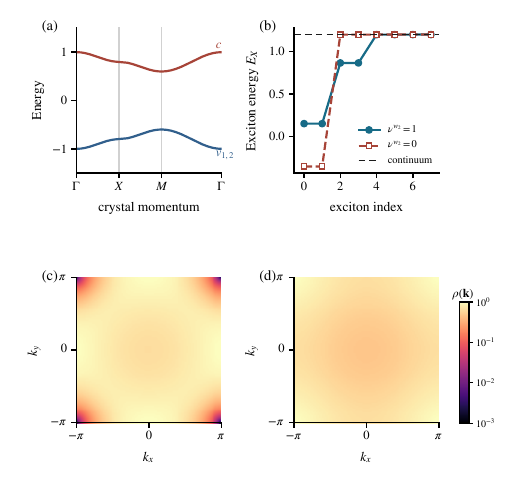}
    \caption{\textit{Two-dimensional $w_2$-obstructed excitons.} (a) Single-particle spectrum of the obstructed and trivial models along $\Gamma\!-\!X\!-\!M\!-\!\Gamma$, consisting of two degenerate valence bands and one trivial conduction band. The minimum direct gap occurs at $M$. (b) Lowest exciton energies for a contact interaction $V(\bq)=U=2$. The dashed line marks the free electron--hole continuum. (c),(d) Momentum-space densities  of the lowest excitons for $\nu^{w_2}=1$ and $\nu^{w_2}=0$, respectively. The densities are normalized to their respective maxima and shown on the same logarithmic scale.}
    \label{fig:w2d2}
\end{figure}
\textit{$w_2$ obstruction in two dimensions.---}In the main text, we consider a three-band model with nontrivial Euler topology. Here, we additionally illustrate a $d=r=2$ SW obstruction using the realification of the Qi--Wu--Zhang (QWZ) model \cite{QiWuZhang2006}, $h_{\mathrm{QWZ}}(\bk)=\mathbf d(\bk)\cdot\boldsymbol{\sigma}$, with
\begin{equation}
\mathbf d(\bk)=\left(\sin k_x,\sin k_y,m+\cos k_x+\cos k_y\right).
\end{equation}
Writing $h_{\mathrm{QWZ}}=A+iB$, its realification is
\begin{equation}
H_R(\bk)=
\begin{pmatrix}
A(\bk)&-B(\bk)\\
B(\bk)&A(\bk)
\end{pmatrix},
\end{equation}
and the rank-two real valence projector is
\begin{equation}
P_v(\bk)=\frac{1}{2}\left[\mathbb I_4-\frac{H_R(\bk)}{|\mathbf d(\bk)|}\right].
\end{equation}
Realification maps the occupied complex QWZ line to an oriented real plane of which its Euler number equals the QWZ Chern number. Consequently, $\nu^{w_2}=\nu^e\bmod 2$. We use $m=1$, which has Chern number $1$, such that $\nu^e=1$ and $\nu^{w_2}=1$. We compare this with the trivial case $m=3$, for which $C=\nu^e=\nu^{w_2}=0$.

The realified QWZ model is used only to define the valence projectors. The active transition space consists of two degenerate valence bands spanning $\operatorname{Im}P_v(\bk)$ and a momentum-independent trivial conduction band, with
\begin{equation}
\varepsilon_{v,a}(\bk)=-\frac{\Delta(\bk)}{2},\qquad \varepsilon_c(\bk)=\frac{\Delta(\bk)}{2},
\end{equation}
where $a=1,2$ and
\begin{equation}
\Delta(\bk)=\Delta_0+t\left(2-\cos k_x-\cos k_y\right).
\end{equation}
We take $\Delta_0=2$ and $t=-0.2$, such that the minimal direct gap at the $M$ point. 

Figure~\ref{fig:w2d2}(a) shows the  single-particle spectrum used for both the trivial and obstructed exciton calculations, while Fig.~\ref{fig:w2d2}(b) shows the corresponding exciton spectra for a contact interaction $V(\bq)=U=2$. The dashed line marks the free electron--hole continuum. Analogous to the Euler excitons in the main text, the lowest $\nu^{w_2}=1$ exciton is less strongly bound than its topologically trivial counterpart. Figures~\ref{fig:w2d2}(c) and \ref{fig:w2d2}(d) show the momentum-space densities of the corresponding lowest excitons. The $\nu^{w_2}=1$ density vanishes at $M$, whereas the $\nu^{w_2}=0$ reference is nodeless. Inversion symmetry pins the zero to $M$ in this model; its existence, however, is enforced by the nontrivial second SW class rather than by inversion symmetry.
\end{document}